\documentclass[natbib]{emulateapj}
\usepackage{natbib}
\input epsf
\def\s2n{S^{\prime}/N}

\usepackage{amssymb,amsmath}
\def\bs{\boldsymbol}

\slugcomment{Submitted to ApJ, \today}
\shorttitle{Supernova Driving. III. Synthetic Molecular Cloud Observations}

\shortauthors{Padoan et al.}

\begin{document}
\title{Supernova Driving. III. Synthetic Molecular Cloud Observations}

\author{Paolo Padoan,}
\affiliation{ICREA \& Institut de Ci\`{e}ncies del Cosmos, Universitat de Barcelona, IEEC-UB, Mart\'{i} Franqu\`{e}s 1, E08028 Barcelona, Spain; ppadoan@icc.ub.edu}
\author{Mika Juvela,}
\affiliation{Department of Physics, PO Box 64, University of Helsinki, 00014, Helsinki, Finland; mika.juvela@helsinki.fi}
\author{Liubin Pan}
\affiliation{Harvard-Smithsonian Center for Astrophysics, 
60 Garden Street, Cambridge, MA 02138, USA; 
lpan@cfa.harvard.edu}
\author{Troels Haugb{\o}lle,}
\affiliation{Centre for Star and Planet Formation, Niels Bohr Institute and Natural History Museum of Denmark, University of Copenhagen, {\O}ster Voldgade 5-7, DK-1350 Copenhagen K, Denmark; haugboel@nbi.ku.dk}
\author{{\AA}ke Nordlund,}
\affiliation{Centre for Star and Planet Formation, Niels Bohr Institute and Natural History Museum of Denmark, University of Copenhagen, {\O}ster Voldgade 5-7, DK-1350 Copenhagen K, Denmark; aake@nbi.ku.dk}

\begin{abstract}

We present a comparison of molecular clouds (MCs) from a simulation of supernova-driven interstellar medium (ISM) turbulence with real MCs 
from the Outer Galaxy Survey. The radiative transfer calculations to compute synthetic CO spectra are carried out assuming the CO relative
abundance depends only on gas density, according to four different models. Synthetic MCs are selected above a threshold brightness 
temperature value, $T_{\rm B,min}=1.4$ K, of the $J=1-0$ $^{12}$CO line, generating 16 synthetic catalogs (four different spatial resolutions and
four CO abundance models), each containing up to several thousands MCs. The comparison with the observations focuses on the 
mass and size distributions and on the velocity-size and mass-size Larson relations. The mass and size distributions are found to be consistent with the 
observations, with no significant variations with spatial resolution or chemical model, except in the case of the unrealistic model with constant 
CO abundance. The velocity-size relation is slightly too steep for some of the models, while the mass-size relation is a bit too shallow for all models
only at a spatial resolution $dx\approx 1$ pc. The normalizations of the Larson relations show a clear dependence on spatial resolution, for both the 
synthetic and the real MCs. The comparison of the velocity-size normalization suggests that the SN rate in the Perseus arm is approximately 70\%
or less of the rate adopted in the simulation. Overall, the realistic properties of the synthetic clouds confirm that supernova-driven turbulence can explain
the origin and dynamics of MCs.

\end{abstract}


\section{Introduction}

The origin, structure and dynamical evolution of molecular clouds (MCs) determine the conversion of gas into stars in galaxies. In the same way 
as modeling of individual prestellar cores cannot form the basis of a star formation theory, the modeling of individual MCs cannot yield a self-consistent 
picture of the cold interstellar medium (ISM). Galactic gas infall, large-scale disk instabilities, spiral arm shocks and supernova (SN) explosions force the 
ISM of galaxies into a highly dynamic state that couples a broad range of scales, challenging the view of MCs as isolated systems. 

The chaotic nature of the ISM dynamics and the unavoidable development of supersonic turbulence result in near-universal statistics and scaling laws, offering
a viable theoretical background to model the fragmentation of the cold ISM leading to star formation. However, a self-consistent theory or numerical model
of the formation and evolution of MCs cannot forgo a physical description of the large-scale forces mentioned above. Although the relative importance of those 
driving mechanisms has not been fully established, and large-scale gravitational instabilities in galactic disks \citep[e.g.][]{Elmegreen+2003,Bournaud+10} and 
gas compression in spiral density waves \citep[e.g.][]{Semenov+15} may also contribute to the turbulence, it is generally accepted that SN explosions dominate 
the energy budget of star-forming galaxies at MC scales \citep[e.g.][]{Ostriker+10sfr,Ostriker+Shetty11,Faucher-Giguere+13,Lehnert+13}.

This is the third of a series of papers where we explore the role of SN driving, based on an adaptive-mesh-refinement (AMR) magneto-hydrodynamic (MHD) 
simulation of a (250 pc)$^3$ ISM volume that could be viewed as a dense section of a Galactic spiral arm. In the first work \citep[][--Paper I hereafter]{Padoan+16mc}, 
we presented 
the simulation and studied the global velocity scaling of SN-driven turbulence, the driving and the scaling of the turbulence within dense clouds, and the resulting
cloud properties in comparison with those of MCs from the Outer Galaxy Survey \citep{Heyer+01}. The large dynamic range of the simulation (it reaches a spatial 
resolution of 0.24 pc), means that, not only individual MCs and SN remnants are well resolved, but they also form self-consistently as the result of larger-scale 
dynamics, with realistic initial and boundary conditions (which is not possible when simulating individual MCs). The simulation provides a large sample of clouds
and thus constrains also the probability distribution of could properties. In Paper I, we found that clouds selected from the simulation
have mass and size distributions, and velocity-size and mass-size relations consistent with the observations. Using tracer particles, we also
studied their evolution and found that they form and disperse in approximately four dynamical times.\footnote{The dynamical time was defined as the ratio of
the cloud radius and the three-dimensional velocity dispersion: $t_{\rm dyn} = R_{\rm cl}/\sigma_{\rm v,3D}$.} 

In the second paper of this series \citep[][--Paper II hereafter]{Pan+16}, we focused on the compressive ratio (the ratio between compressive and solenoidal modes) of the 
turbulence within individual MCs selected from our simulation and its relation to the statistics of density fluctuations \citep[e.g.][]{Federrath+08,Schmidt+09,
Kritsuk+10,Federrath+10,Kritsuk+11codes}, which is a main ingredient in models of the star formation rate \citep[e.g.][]{Padoan95,Krumholz+McKee05sfr,
Padoan+Nordlund11sfr,Hennebelle+Chabrier11sfr,Federrath+Klessen12} and of the stellar initial mass function \citep[e.g.][]{Padoan+97imf,
Padoan+Nordlund02imf,Hennebelle+Chabrier08,Hopkins12imf}. In Paper II, we showed that SN-driven turbulence results in a broad range of values of the compressive 
ratio in MCs, with an approximately lognormal probability distribution with a peak at $\approx 0.3$. The cloud self-gravity does not affect this distribution
significantly, though this remains to be verified with longer simulations including self-consistent SN feedback (resolving the formation of massive stars that 
will later explode as SNe), and with simulations with larger total column density (or lower SN rate), possibly generating clouds with lower virial parameters. 

In this third study, we further test the consistency of MC properties from our simulation with those of observed MCs. As in Paper I, simulated clouds
are compared with those from the Outer Galaxy Survey \citep{Heyer+01}. In Paper I, we considered projected properties (e.g. line-of-sight
velocity, equivalent radius) of clouds selected in three spatial dimensions from the computational volume; here, we solve for the radiative transfer through
the simulation volume and compute synthetic spectra of CO emission lines that are analyzed to select MCs with the same criteria as in the observations.   

The simulation is briefly presented in the next section, while in section~\ref{sect_rt} we describe our assumed CO abundance models and the radiative transfer 
calculations. In section~\ref{sect_mcs} we describe the cloud selection, following \citet{Heyer+01}, and the generation of 16 synthetic MC catalogs. In the following
two sections we compare the mass and size distributions and the Larson relations of the observations with those from our fiducial catalog, while results from the 
other catalogs (with different CO abundance models or spatial resolutions) are discussed in section~\ref{sect_resol}. Our conclusions are summarized in 
section~\ref{sect_conclusions}.

\section{Simulation}

This work is based on an MHD simulation of SN-driven turbulence in a large ISM volume, carried out with the Ramses AMR code \citep{Teyssier02}. 
The numerical method and setup are discussed extensively in Paper I and are only briefly summarized here. We simulate a cubic region 
of size $L_{\rm box}=250$ pc, with a minimum cell size of $dx=0.24$ pc, periodic boundary conditions,
a mean density of 5 cm$^{-3}$ (corresponding to a total mass of $1.9\times 10^6$ $M_{\odot}$) and a core-collapse SN rate of 6.25 Myr$^{-1}$. 
We distribute SN explosions randomly in space and time (see discussion in Paper I in support of this choice), so our SN 
rate could also be interpreted as the sum of all types of SN explosions. Individual SN explosions are implemented with an instantaneous 
addition of 10$^{51}$ erg of thermal energy and 15 M$_{\odot}$ of gas, distributed according to an exponential profile on a spherical 
region of radius $r_{\rm SN}=3 dx=0.73$ pc, which guarantees numerical convergence of the SN remnant evolution \citep{Kim+Ostriker15SN}. 

Our total energy equation includes the $p dV$ work, the thermal energy introduced to model SN explosions, 
uniform photoelectric heating up to a critical density of 200 cm$^{-3}$, and parametrized cooling functions from \citet{Gnedin+Hollon12}. 
The simulation is started with zero velocity, a uniform density $n_{\rm H,0}=5$ cm$^{-3}$, a uniform magnetic field $B_0=4.6$ $\mu$G (chosen 
to achieve near equipartition with the kinetic energy at large scales) and a uniform temperature $T_0=10^4$ K. The first few SN explosions rapidly 
bring the mean thermal, magnetic and kinetic energy to approximately steady-state values, with the magnetic field amplified to an rms value of 7.2 $\mu$G
and an average of $|\bs B|$ of 6.0 $\mu$G, consistent with the value of $6.0 \pm 1.8$ $\mu$G derived from the `Millennium Arecibo 21-cm Absorption-Line 
Survey' by \citet{Heiles+Troland05}.We have run the simulation for 45 Myr without self-gravity and then continued with self-gravity for 11 Myr. 

Previous computational work on SN-driven turbulence was briefly reviewed in Papers I and II. A concise review of the variety of computational 
setups used to study ISM turbulence can also be found in \cite{Kritsuk+11}, where the formation of MCs as a result of large-scale ISM turbulence
is outlined based on simulations of a 200 pc region of the multi-phase ISM driven by a random external force, instead of SN explosions.

\section{Radiative Transfer Calculation} \label{sect_rt}

\begin{figure}[t]
\includegraphics[width=\columnwidth]{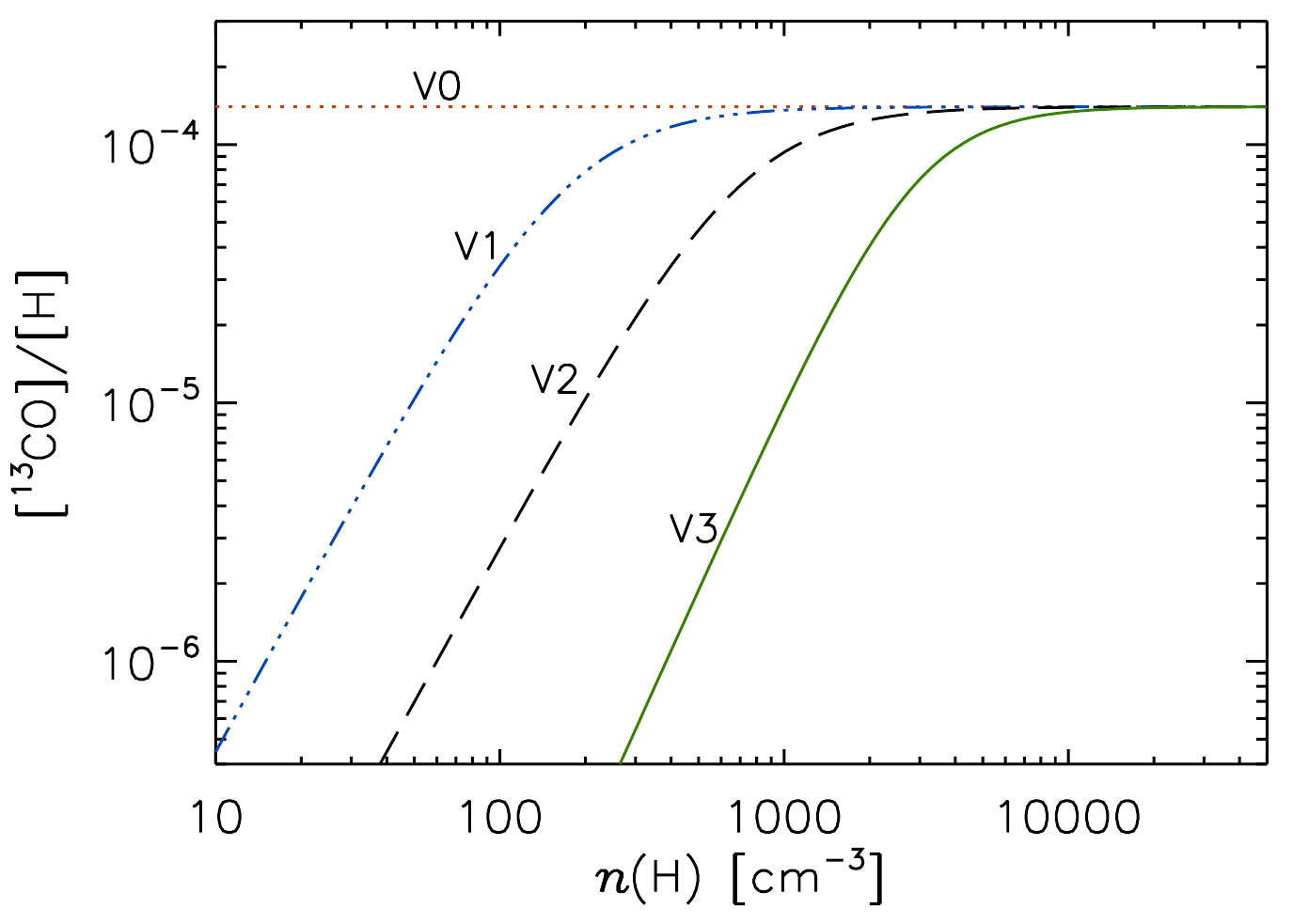}
\caption[]{Relative abundance of CO versus gas density in the four models adopted in this work. The versions V2 and V3 correspond to the two main
groups of model predictions shown in \citet{Glover+Clark12} (see their Fig. 3a).}
\label{abundance}
\end{figure}

For calculations of synthetic CO spectra, the density and velocity data are resampled onto a 512$^3$ grid. 
Because the simulation does not include the computation of chemical networks, the CO abundance must be modeled
based on results of previous studies. We adopt a simple model where the CO relative abundance, [CO]/[H$_2$], depends only 
on the local gas density, and test four different versions of such dependence:
\begin{align*} 
{\rm [CO]/[H_2]}  & = A  & {\rm(V0)} \\
&  = A \,16 \,n_{\rm H}^{2}  / (5\times 10^{5} + 16 \,n_{\rm H}^{2})  & {\rm(V1)}\\
&  = A \,  n_{\rm H}^{2 } / (5\times 10^{5} + n_{\rm H}^{2})  & {\rm(V2)}  \\
&  = A \,  n_{\rm H}^{2.45} / (3.8\times 10^{8} + n_{\rm H}^{2.45})  & {\rm(V3)}
\end{align*}
where $A=1.4\times 10^{-4}$ and $n_{\rm H}$ is in cm$^{-3}$.
These models are shown in Fig.~\ref{abundance}. The density threshold for the
appearance of CO molecules systematically increases from version V0 to version V3. The
version V0 has a constant abundance; it is not a realistic model, but we include it as a reference. 
The versions V2 and V3 correspond to the two main
groups of model predictions shown in \citet{Glover+Clark12} (see their Fig. 3a). Those model predictions
are results of simulations of chemistry in turbulent clouds with a mean density of
$n=$100\,cm$^{-3}$. The higher density threshold (thus V3) is preferred on the grounds
that it corresponds to more complete models of carbon chemistry. On the other hand, in
clouds of higher mean density, CO can appear already at lower local densities.
Therefore, we include the abundance profile V1 to fill in the gap between the constant
abundance case and the version V2. In all cases, at high densities the relative abundance approaches
the same chosen value of [CO]/[H$_2$]=$1.4 \times 10^{-4}$, corresponding to the assumed 
total elemental abundance of carbon \citep{Glover+Clark12}. Considering the
expense of full chemistry calculation, a pure density dependence is a convenient way to
take into account the first order effects of abundance variations. However, it is only
an approximation and in reality the relation between abundance and density contains a 
significant scatter \citep[see e.g.][]{Glover+10}.

The radiative transfer calculations are carried out using the new line transfer program
LOC (Juvela, in prep.) that uses OpenCL libraries to parallelize the calculations. To
speed up the convergence of optically thick lines, the program also uses accelerated
lambda iterations \citep{Olson+86}, similar to the implementation in
\citet{Juvela+Padoan05}. The program has been tested against the Cppsimu Monte Carlo code
\citep{Juvela97}, which was used to repeat the calculations for a few of the models.
LOC simulates the radiation field with the ray-tracing method, the present calculations
employing rays with 48 different directions.  The calculations cover a velocity
bandwidth of 49.6\,km\,s$^{-1}$ with 0.202\,km\,s$^{-1}$ channels. The channel width is chosen
so that the average of four channels exactly corresponds to the 0.81\,km\,s$^{-1}$
channel width used in the cloud extraction. The simulation of the radiation field and
the updates of the level populations are iterated until convergence. In particular, the
line intensity maps are monitored to make sure that the changes, even over several
consecutive iterations, have decreased to a level of 1\% or below. Maps of 512$\times$512
spectra, in units of brightness temperature, $T_{\rm B}= \lambda^2/(2k) I_{\nu}$, are calculated 
for the three orthogonal coordinate directions, one map pixel thus corresponding to a linear scale of 0.49 pc. 

Four molecular transitions are computed, $J=1-0$ and $J=2-1$ of $^{12}$CO and $^{13}$CO.
However, here we use only the $^{12}$CO $J=1-0$ line, as our main purpose is the comparison
with observations of the same line from the MCs in the Outer Galaxy Survey \citep{Heyer+01}.
Results from the other lines will be discussed elsewhere.

\section{Molecular Cloud Selection} \label{sect_mcs}

As in Paper I, the comparison with the observations is based on the MC catalog by \citet{Heyer+01}, 
extracted from the $^{12}$CO Five College Radio Astronomy Observatory (FCRAO) Outer Galaxy 
Survey \citep{Heyer+98}. With a total number of 10156 objects, up to a mass of approximately $8\times 10^5$ $M_{\odot}$ 
and a size of 45 pc, this is the largest sample of Galactic MCs available. It is estimated to be complete down to a mass of 
approximately 600 M$_{\odot}$ and a cloud size of 3 pc, as velocity blending is not as serious as in the case of Inner-Galaxy 
surveys.

As in \citet{Heyer+01}, we consider only the subset of 3,901 clouds with circular velocities $v_{\rm c} < -20$ km s$^{-1}$, 
because of kinematic distance accuracy. Given the distances to these clouds and the angular resolution of the survey, the 
spatial resolution varies between 0.4 pc and 3.8 pc. As discussed below (see section~\ref{sect_resol}), we find that the normalization 
of the velocity-size relation (to a lower extent also the slope) depends on the spatial resolution. In order to make the comparison
between the simulation and the observation more specific, we have thus divided the observational sample into relatively narrow
distance (spatial resolution) intervals, as described in section~\ref{sect_resol}.

We select MCs from the synthetic spectra following as closely as possible the steps described in \citet{Heyer+01}. For the fiducial model
presented in the next section, we adopt the version V2 of the CO abundance model, and resample the grid of $512\times512$ synthetic 
spectra into a $256\times256$ grid, resulting in a spatial resolution of 0.98 pc, as that is the resolution corresponding to our distance
interval with the largest number of MCs from the Outer Galaxy Survey. We then average over 4 velocity channels to exactly match the 
0.81\,km\,s$^{-1}$ channel width of the observations, and add Gaussian noise to mimic the median rms temperature of 0.93 K in channels 
with no emission reported by \citet{Heyer+98}. MCs are then selected as connected regions in position-position-velocity (PPV) space above
a threshold brightness temperature value $T_{\rm B,min}=1.4$ K\footnote{In \citet{Heyer+01}, the threshold value of 1.4 K is for the main 
beam brightness temperature (the antenna temperature corrected for the main beam efficiency of the telescope), and the corresponding 
value in brightness temperature may be slightly different, depending on the source brightness distribution within the beam.}, satisfying the 
following conditions: i) at each spatial position, at least two contiguous velocity channels must be above the threshold to be selected (single 
pixels above the threshold, isolated in velocity, are discarded); ii) a minimum of 5 spatial positions are required for an object to be included 
(thus a minimum of 10 PPV pixels, given the first condition). We apply this selection method to data cubes from 3 simulation snapshots at 7, 9 
and 11 Myr after gravity was included in the simulation (the third snapshot is the last one of the run). Because each snapshot is used for three 
grids of synthetic spectra (one in each coordinate direction), nine PPV data cubes are generated, yielding a total of 1250 synthetic MCs in our 
fiducial catalog (chemical abundance model V2 and 0.98 pc resolution).

\begin{table}[t]\footnotesize
\caption{Properties of the 16 synthetic MC catalogs and the 
six distance intervals of the observational catalog.}
\centering
\begin{tabular}{ccrrrrc}
\hline\hline \\[-2.2ex]
\multicolumn{7}{c}{Synthetic Catalogs:}        \\
\hline
CO/H$_2$  & $dx$ & $N_{\rm cl}$ & $M_{\rm tot}$ & $M_{\rm min}$  &  $M_{\rm max}$ & $\Sigma_{\rm aver}$     \\ 
                   & [pc]   &                      & [$M_{\odot}$] & [$M_{\odot}$]    &  [$M_{\odot}$]    & [$M_{\odot}$pc$^{-2}$]   
\\ [0.8ex]
\hline  
V0    & 0.49 &  19017 &  1.0e7   &   1.5       &  2.0e6  & 18.3  \\
V0    & 0.98 &  4933   &  1.0e7 &   23.6     &  1.9e6    & 19.2  \\
V0    & 1.95 &  1271   &  1.0e7 &   129.2   &  2.0e6    & 20.3  \\
V0    & 3.91 &  317     &  0.9e7   &   685.3   &  1.7e6  & 22.3  \\
\hline 
V1    & 0.49 &  6352  &  5.0e6   &   2.4        &  7.2e5  &  21.7  \\
V1    & 0.98 &  2158  &  4.6e6   &   36.0      &  6.4e5  &  23.3  \\
V1    & 1.95 &  612    &  4.1e6   &   153.1    &  3.7e5  &  25.0  \\
V1    & 3.91 &  183    &  3.4e6   &   1009.0  &  2.7e5  &  26.4  \\
\hline 
V2    & 0.49 &  4021  &  1.5e6   &   7.4       &  8.1e4   &  23.3  \\
{\bf V2}    & {\bf 0.98} &  {\bf 1250}  &  {\bf 1.4e6}   &   {\bf 35.2}     &  {\bf 7.9e4}   &  {\bf 24.2}  \\
V2    & 1.95 &  348    &  1.1e6   &   174.8   &  8.3e4   &  24.6  \\
V2    & 3.91 &  85      &  0.8e6   &   732.0   &  6.3e4   &  26.0  \\
\hline 
V3    & 0.49 &  1565 &  3.0e5    &   7.0       &  1.2e4   &  27.6  \\
V3    & 0.98 &  438   &  2.6e5    &   45.5     &  1.3e4   &  29.0  \\
V3    & 1.95 &  109   &  1.6e5    &   181.2   &  1.0e4   &  30.8  \\
V3    & 3.91 &  15     &  0.5e5    &   1055.4 &  0.7e4   &  24.0  \\     
\hline \\[-2.2ex]
\multicolumn{7}{c}{Observational Catalogs:}        \\
\hline
$D_{\rm aver}$  & $dx$ & $N_{\rm cl}$ & $M_{\rm tot}$ & $M_{\rm min}$  &  $M_{\rm max}$ & $\Sigma_{\rm aver}$   \\  
      $\rm [kpc]$  & [pc]   &                      & [$M_{\odot}$] & [$M_{\odot}$]    &  [$M_{\odot}$]    & [$M_{\odot}$pc$^{-2}$]   
\\ [0.8ex]
\hline     
2.0   &  0.50  &  212  &  8.1e4  &  11.1  &  3.9e4  &  18.0 \\    
3.1   &  0.78  &  584  &  4.0e5  &  23.0  &  6.2e4  &  21.0 \\
{\bf 4.1}   &  {\bf 1.04}  & {\bf 1936} &  {\bf 3.1e6}  &  {\bf 36.9}  &  {\bf 3.0e5}  &  {\bf 21.7} \\
5.4   &  1.36  & 1816 &  4.0e6  &  49.2  &  8.2e5  &  19.2 \\
7.5   &  1.89  &  360  &  5.1e5  & 131.2 &  3.3e4  &  15.8 \\
10.2 &  2.55  &  88    &  1.4e5  & 319.8 &  1.9e4  &  14.2 \\
\hline
\end{tabular}
\label{t1}
\end{table}

\begin{figure}[t]
\includegraphics[width=\columnwidth]{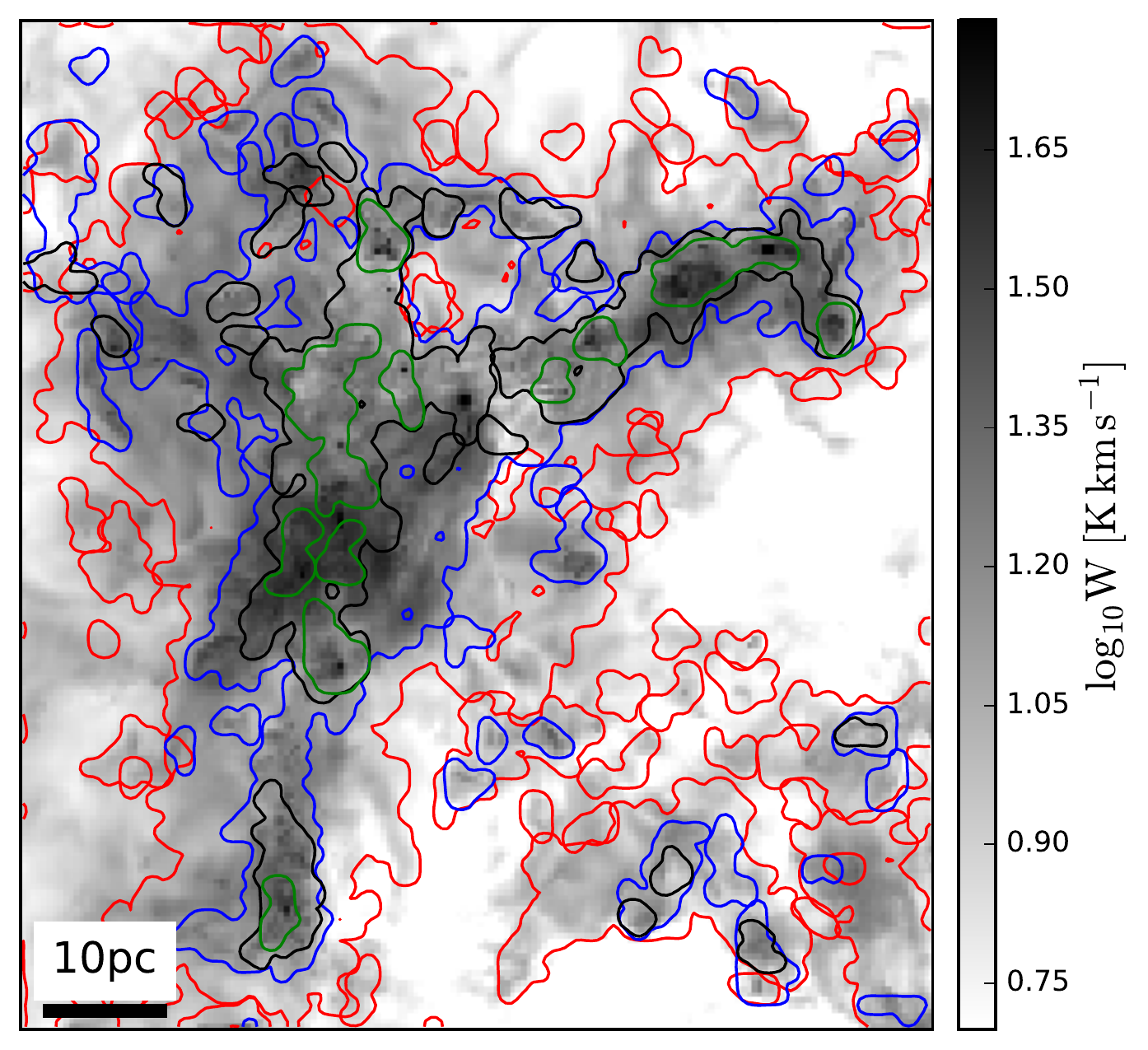}
\caption[]{Example of clouds extracted from the last snapshot of the simulation, in the area around the most massive MC. The background image 
is the line area from model V0 with 0.49 pc resolution. The red, blue, black and green contours show the outlines of all clouds extracted in this area 
from models V0, V1, V2, and V3, respectively, at a resolution of 0.98 pc.}
\label{images}
\end{figure}

Besides this fiducial catalog, we create three more using the same chemical abundance model and three different spatial resolutions, 
using the original grid of $512\times512$ synthetic spectra, or resampling it into $64\times64$ and $128\times128$ spectra. We also 
compute four catalogs of four different spatial resolutions for each of the other three abundance models, V0, V1 and V3, for a total of
16 synthetic MC catalogs. The number of MCs, their total mass, the minimum and maximum cloud mass, and 
the average cloud surface density for each of the 16 catalogs (and for the six observational ones) are given in Table \ref{t1}. 
The table shows that the total cloud number and
mass, as well as the maximum cloud mass decrease as the CO abundance decreases from model V0 through V3. On the contrary, the 
cloud mean surface density slightly increases, as cloud contours gradually move towards inner higher-density regions with decreasing CO 
abundance (models V0 through V3). Examples of cloud contours from the four chemical models are shown in Figure \ref{images}. 
 
In our simulation, clouds of a wide range of masses arise as density fluctuations in the turbulent ISM, where the dynamics is approximately 
scale free down to the scale of collapsing prestellar cores, where gravity is dominant. Thus, as in Paper I, despite the 
mass range of up to six orders of magnitude, we refer to all clouds as MCs, instead of following the usual classification into clumps, 
molecular clouds, giant molecular clouds, and giant molecular cloud complexes, in order of increasing mass. 

In the following two sections, we derive the mass and size distributions and the Larson relations of our synthetic MCs and compare them 
with the observations. We only consider the fiducial catalog (chemical abundance model V2 and 0.98 pc resolution -- upper boldface row in Table \ref{t1}) 
and the sub-sample of the observational survey within the corresponding distance interval ($\sim 4$ kpc corresponding to $\sim 1$ pc resolution -- lower 
boldface row in Table \ref{t1}) of the observational survey. The other synthetic catalogs (and distance intervals of the observations) are 
discussed in section~\ref{sect_resol}.

\section{Mass and Size Distributions} \label{sect_mass_size}

Cloud masses are derived with equation (6) of \citet{Heyer+01}, which assumes a constant CO-to-H$_2$ conversion factor:
\begin{equation}
M_{\rm cl}=4.1 \left( {L_{\rm CO} \over {\rm K\,km\,s^{-1}pc^2}}\right)\, {M}_{\odot},
\label{eq_mass}
\end{equation}
where $L_{\rm CO}$ is the CO luminosity in units of ${\rm K\,km\,s^{-1}pc^2}$, defined as the sum of the brightness temperature, $T_{\rm B}(x,y,v)$,
over all PPV cells contained in a cloud,
\begin{equation}
L_{\rm CO}=\sum_{\rm i,j,k} T_{\rm B}(x_{\rm i},y_{\rm j},v_{\rm k}) dx^2 dv,
\label{eq_Lco}
\end{equation}
where $dx$ is the spatial cell size in pc ($dx = 0.98$ pc in our fiducial catalog), $dv$ is the velocity channel width
in km~s$^{-1}$ ($dv=0.81$ km~s$^{-1}$ in all catalogs), and the brightness temperature is the output of the radiative transfer calculation
described in the previous section.

\begin{figure}[t]
\includegraphics[width=\columnwidth]{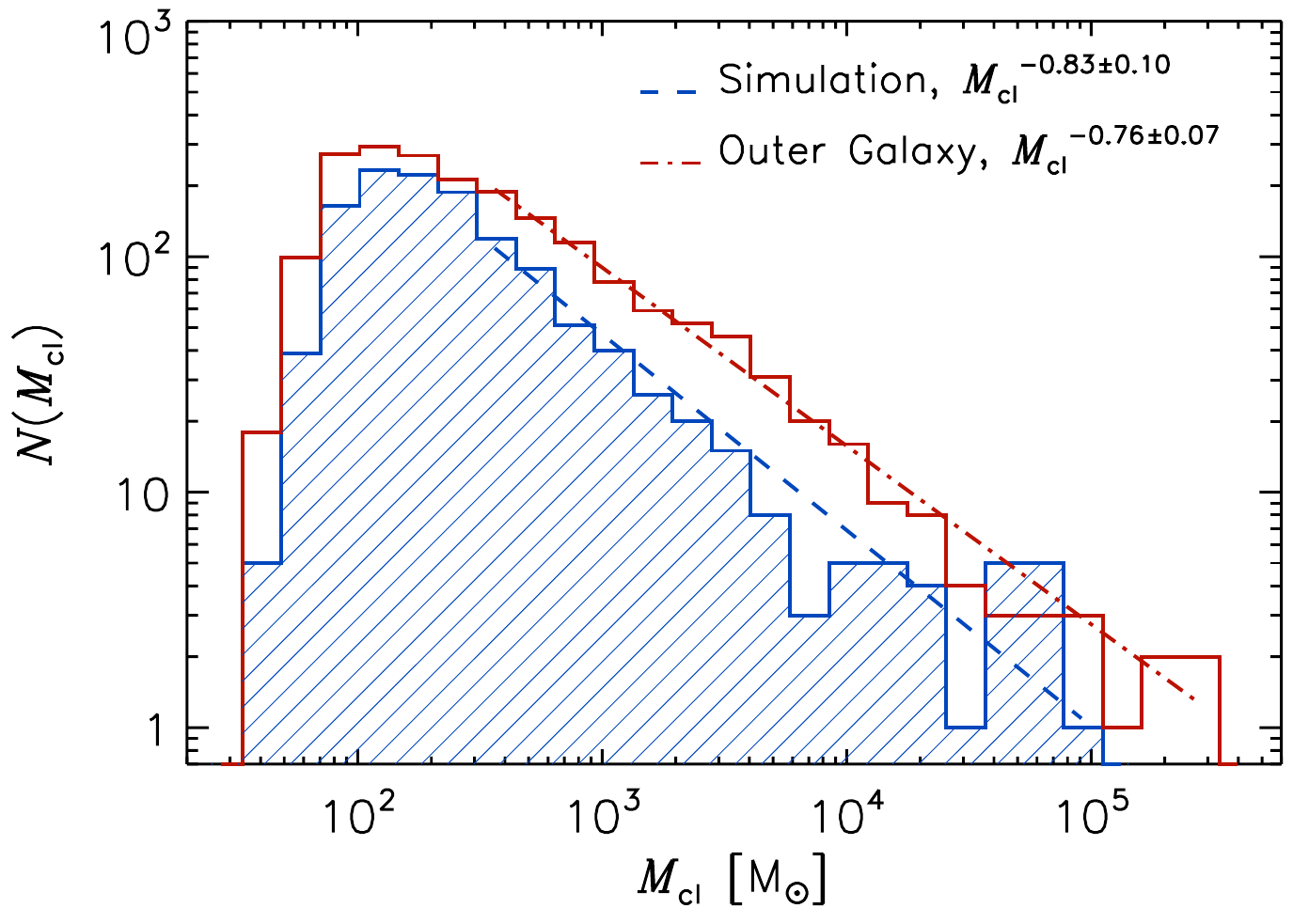}
\caption[]{Probability distribution of cloud masses for the sample of 1250 clouds from our fiducial synthetic catalog, based on the chemical-abundance 
model V2 and spatial resolution $dx=0.98$ pc (shaded histogram). The unshaded histogram shows the mass distribution from the observational 
sample of 1936 Outer-Galaxy MCs in the distance interval $[4/1.25,4\times1.25]$ kpc. The power-law fits to the distributions, shown by the dashed and 
dashed-dotted lines, are evaluated for masses above 300 M$_{\odot}$, just above the estimated completeness limit of the Outer Galaxy Survey at the 
distance of 4 kpc.}
\label{mass}
\end{figure}

The cloud radius is defined as the radius of the circle with area equal to the cloud projected area:
\begin{equation}
R_{\rm e}=\sqrt{{1 \over \pi}\sum_{\rm i,j} dx^2}.
\label{eq_Lco}
\end{equation}
Given the conditions for cloud selection explained in the previous section, MCs from our fiducial catalog must have a mass larger than 
10.9 $M_{\odot}$ and a radius larger than 0.55 pc. As shown in Table \ref{t1} (boldface row), the actual minimum mass is 35.2 $M_{\odot}$.

Before comparing with the observations, we add a random uncertainty to the derived values of masses and radii, corresponding to a statistical uncertainty in distance of 10\%, probably somewhat smaller than for most of the observed clouds (the distance uncertainties for the clouds in the Outer Galaxy Survey are not given in \citet{Heyer+01}).

\begin{figure}[t]
\includegraphics[width=\columnwidth]{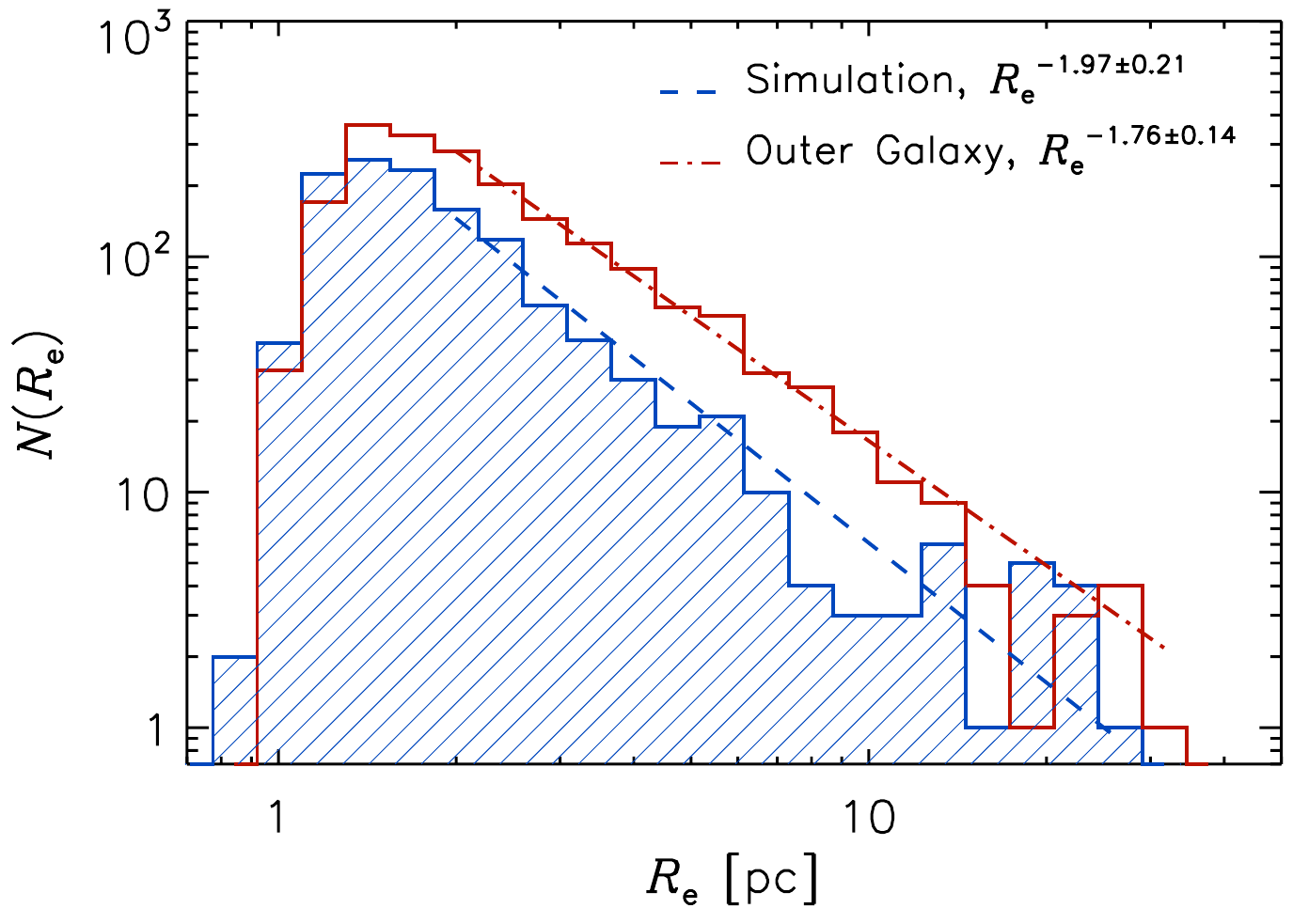}
\caption[]{Probability distribution of cloud equivalent radii for the same fiducial synthetic cloud catalog (shaded histogram) and the same observational 
sample (unshaded histogram) as in Figure \ref{mass}. The power-law fits to the distributions (dashed and dashed-dotted lines) are evaluated for radii 
above 2 pc, just above the estimated completeness limit of the Outer Galaxy Survey at the distance of 4 kpc.}
\label{size}
\end{figure}

The mass and size probability distributions of the synthetic clouds from our fiducial catalog are shown in Figures \ref{mass} and \ref{size}, together
with those from the observational sample limited to the 1936 clouds in the distance interval $[4/1.25,4\times1.25]$ kpc, to match the spatial resolution
of approximately 1 pc of the synthetic catalog. The mass and radius completeness limits of the Outer Galaxy Survey estimated by \citet{Heyer+01} for a 
distance of 10 kpc are 600 M$_{\odot}$ and 3.1 pc respectively. Because here we consider only a subsample of clouds at a distance around 4 kpc,
we may expect the completeness limit to be lower by a factor of 4 and 2 respectively. Because of the finite width of our distance interval, we adopt a slightly 
more conservative mass and size limit, only 2 and $\sqrt{2}$ times smaller than those of the full sample. Thus, we compute power-law fits for clouds 
more massive than approximately 300 M$_{\odot}$ and larger than approximately 2 pc. The same limits are adopted for the fiducial catalog as well. 
We find that the slopes of the distributions from the synthetic catalog are consistent with the observations within the 1-$\sigma$ uncertainty. 
Expressing the mass distribution as $N(M_{\rm cl})\propto M_{\rm cl}^{-\beta_{\rm M}}$ and the size distribution as 
$N(R_{\rm e})\propto R_{\rm e}^{-\beta_{\rm R}}$, we find $\beta_{\rm M}=0.83\pm0.10$ and $0.76\pm0.07$ and $\beta_{\rm R}=1.97\pm0.21$ and 
$1.76\pm0.14$ for the simulation and the observations respectively.

\section{Larson Relations} \label{sect_larson}

\begin{figure}[t]
\includegraphics[width=\columnwidth]{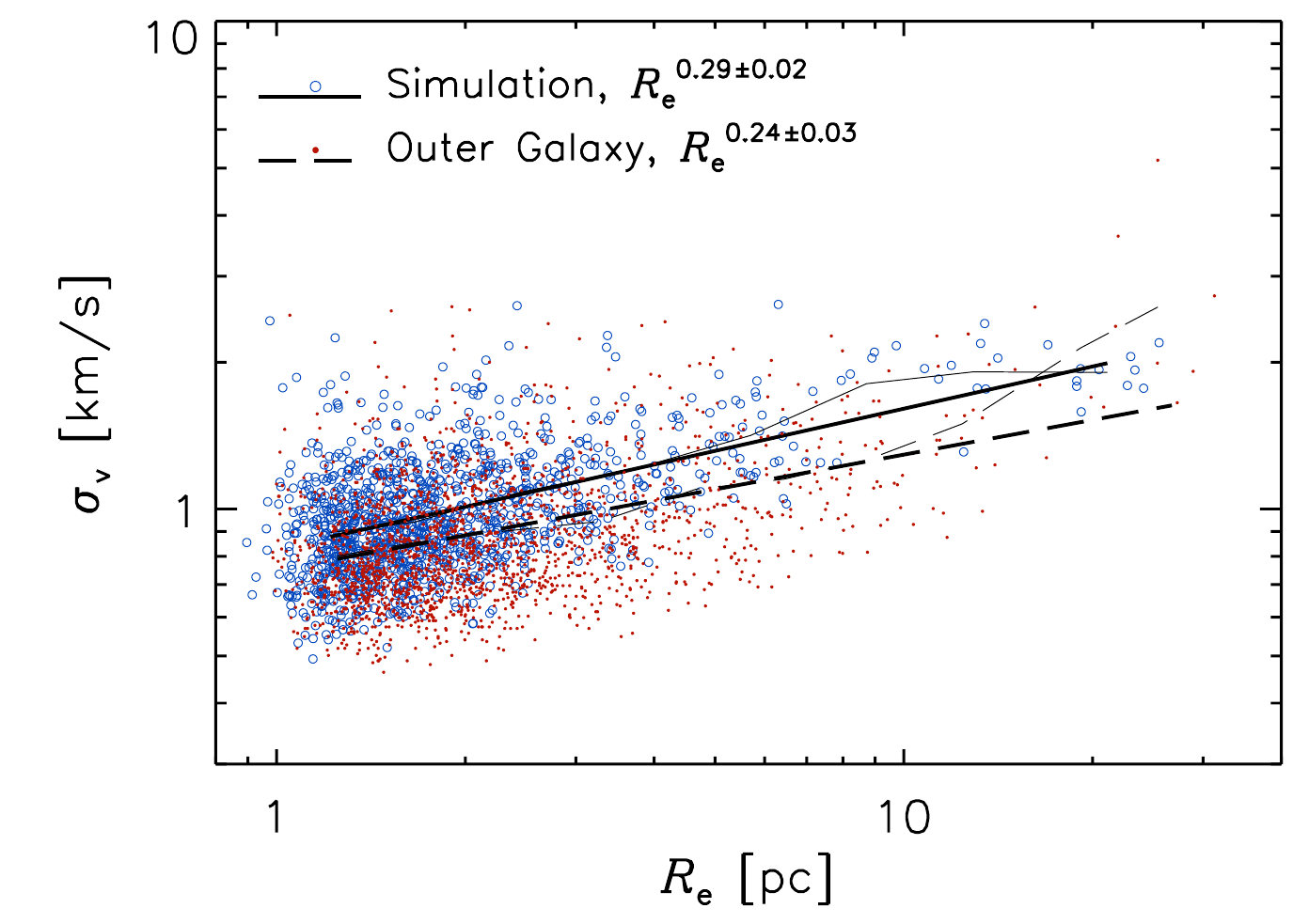}
\caption[]{Velocity-size relation for the MCs selected from the fiducial synthetic catalog (blue empty circles) and from the same observational 
sample as in Figure \ref{mass} (red dots). The values of $\sigma_{\rm v}$ averaged within logarithmic intervals of $R_{\rm e}$ are shown by 
the thin solid and dashed lines. The thick solid and dashed lines show the power-law fits through those average values. Both the slope and 
the scatter in the relations from the two samples are very similar. However, the velocity normalization of the synthetic MCs is approximately 
10\% larger than from the observations.}
\label{velocity_size}
\end{figure}

The line-of-sight rms velocity of a cloud, $\sigma_{\rm v}$, is computed from the equivalent width of the cloud composite spectrum, as in \citet{Heyer+01}:
\begin{equation}
\sigma_{\rm v}={1\over{\sqrt{8\ln{2}}}}\sum_{\rm k} {\psi(v_{\rm k}) dv \over{\max{\psi(v_{\rm k})}}},
\label{eq_sigmav}
\end{equation}
where $\psi(v_{\rm k})$ is the composite spectrum,
\begin{equation}
\psi(v_{\rm k})=\sum_{\rm i,j} T_{\rm B}(x_{\rm i},y_{\rm j},v_{\rm k}).
\label{eq_psi}
\end{equation}
The relation between $\sigma_{\rm v}$ and $R_{\rm e}$ is shown in Figure \ref{velocity_size} for both the synthetic catalog (empty circles) and the 
observations (dots). To derive a power-law fit, we bin the data in logarithmic intervals of $R_{\rm e}$, and compute the mean and standard deviation 
of $\sigma_{\rm v}$ in each interval (thin solid and dashed lines in Figure \ref{velocity_size}). We then perform a least-square fit through these mean 
values yielding,
\begin{equation}
\sigma_{\rm v}=\sigma_{\rm v,1pc}\left({R_{\rm e}\over{1\,{\rm pc}}}\right)^{\beta_{\rm v}}
\label{eq_larson}
\end{equation}
with $\sigma_{\rm v,1pc}=0.83\pm 0.02$ km\,s$^{-1}$ and $\beta_{\rm v}=0.29\pm 0.02$ for the synthetic clouds, and $\sigma_{\rm v,1pc}=0.75\pm 0.02$ 
km\,s$^{-1}$ and $\beta_{\rm v}=0.24\pm 0.03$ for the observed clouds.

The 2-$\sigma$ discrepancy between the velocity normalizations is not surprising,
because the simulation was not tailored to reproduce precisely the conditions in the Perseus arm, where most of the observed MCs are located. The SN
rate adopted in the simulation is relatively large, and could easily exceed the actual SN rate in the Perseus arm, causing a slight excess in the velocity 
dispersion of the clouds in the simulations. The turbulence dissipation rate scales as $v^3/\ell$, where $v$ is the velocity dispersion at the scale $\ell$. 
Assuming an equilibrium between turbulent energy dissipation and SN energy injection rates, our SN rate should be reduced to approximately 70\% of 
its current value to recover the observed velocity normalization. Although this lower SN rate remains reasonable, further studies of the star formation rate 
in the Perseus arm are needed to properly constrain the SN rate. The above assumption of equilibrium between SN energy injection and turbulent 
dissipation is non-trivial for a system with diverse cooling and heating mechanisms. However, our conclusion holds even if the two rates are different, as 
long as their ratio is constant (e.g. if the efficiency of the conversion of SN energy into turbulent motions does not strongly depend on the SN rate). 
Furthermore, numerical simulations of SN driven turbulence result in a gas velocity dispersion approximately proportional to the cubic root of the SN 
rate \citep[][Figures 12 and 13]{Dib+06SN}, as assumed in our argument.

\begin{figure}[t]
\includegraphics[width=\columnwidth]{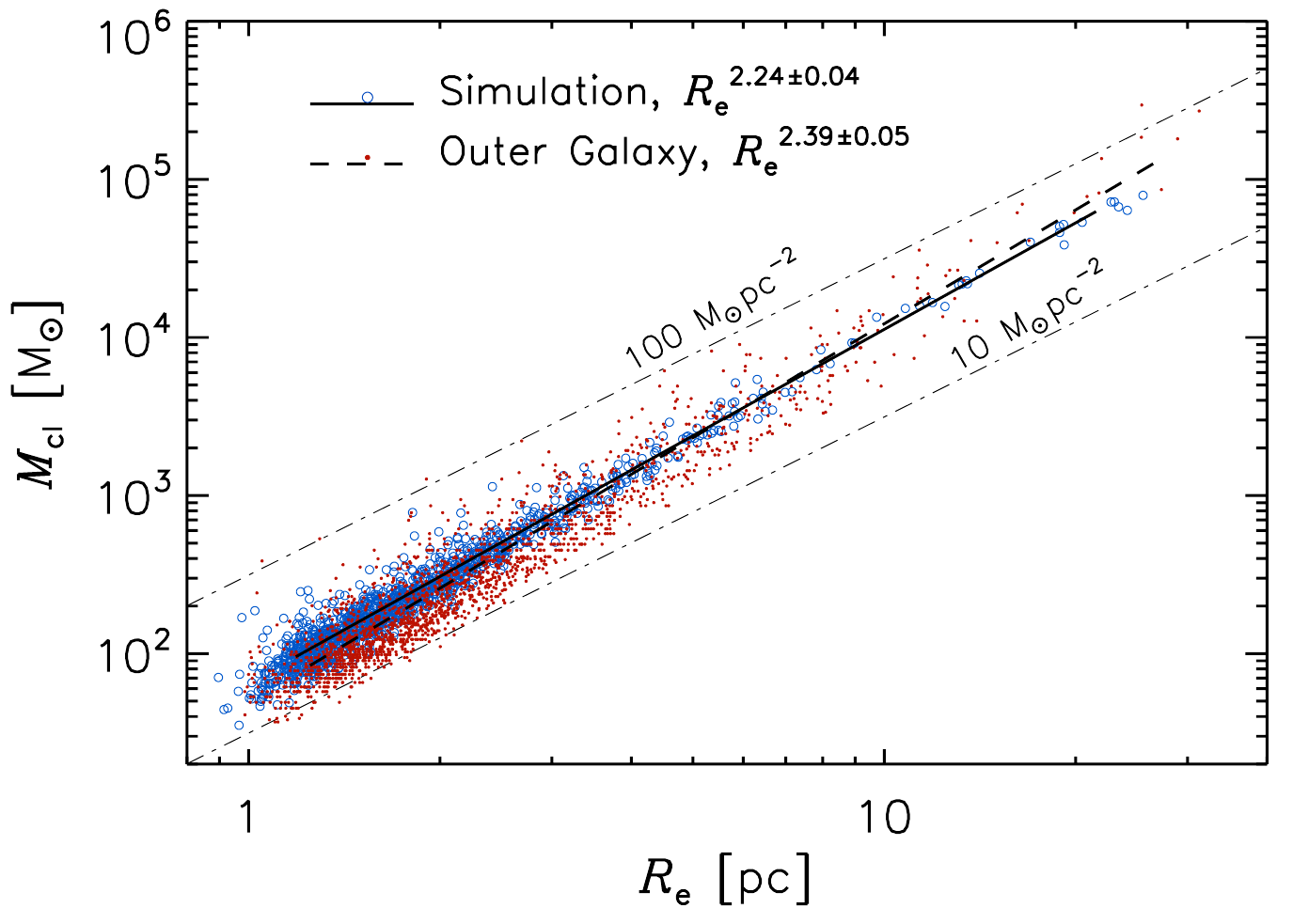}
\caption[]{Mass-size relation for the same synthetic and observational cloud catalogs as in the previous figures. The solid and dashed lines show the
power-law fits to the binned data, as in Figure \ref{velocity_size}. The dashed-dotted lines mark two values of constant surface density.}
\label{mass_size}
\end{figure}

Figure 6 in \citet{Heyer+01} shows that the velocity-size relation derived for the Outer-Galaxy MCs continuously flattens with decreasing values of $R_{\rm e}$,
and is very flat for $R_{\rm e}\lesssim 10$ pc. This flattening is likely due to the inclusion of clouds at different distances in the sample, combined with the 
dependence of the velocity normalization on spatial resolution discussed in the next section (and with the limited velocity resolution). Because we 
select a sub-sample of clouds at approximately the same distance (4 kpc for this comparison with our fiducial synthetic catalog), the flattening in the 
velocity-size relation of the observed MCs is nearly completely eliminated in our case. Our Figure \ref{velocity_size} shows that for this observational 
sub-sample the velocity-size relation is well approximated by a power law, at least for $R_{\rm e}\lesssim 10$ pc. 

We should stress that the slope of the velocity-size relation derived here is affected by the limited velocity resolution. Using the full velocity resolution
of our synthetic spectra we derive a significant number of small clouds with smaller velocity dispersion than in Figure \ref{velocity_size}. Although we
have largely eliminated the flattening of the lower envelope of the observational velocity-size relation by selecting clouds within a relatively narrow 
distance range, it is very likely that, with smaller velocity channels, even the observational survey would result in a large number of small clouds with 
lower velocity dispersion than with the current velocity resolution. The real slope of the velocity-size relation is thus somewhat larger than the values of 
$\beta_{\rm v}$ derived here.

\begin{figure}[t]
\includegraphics[width=\columnwidth]{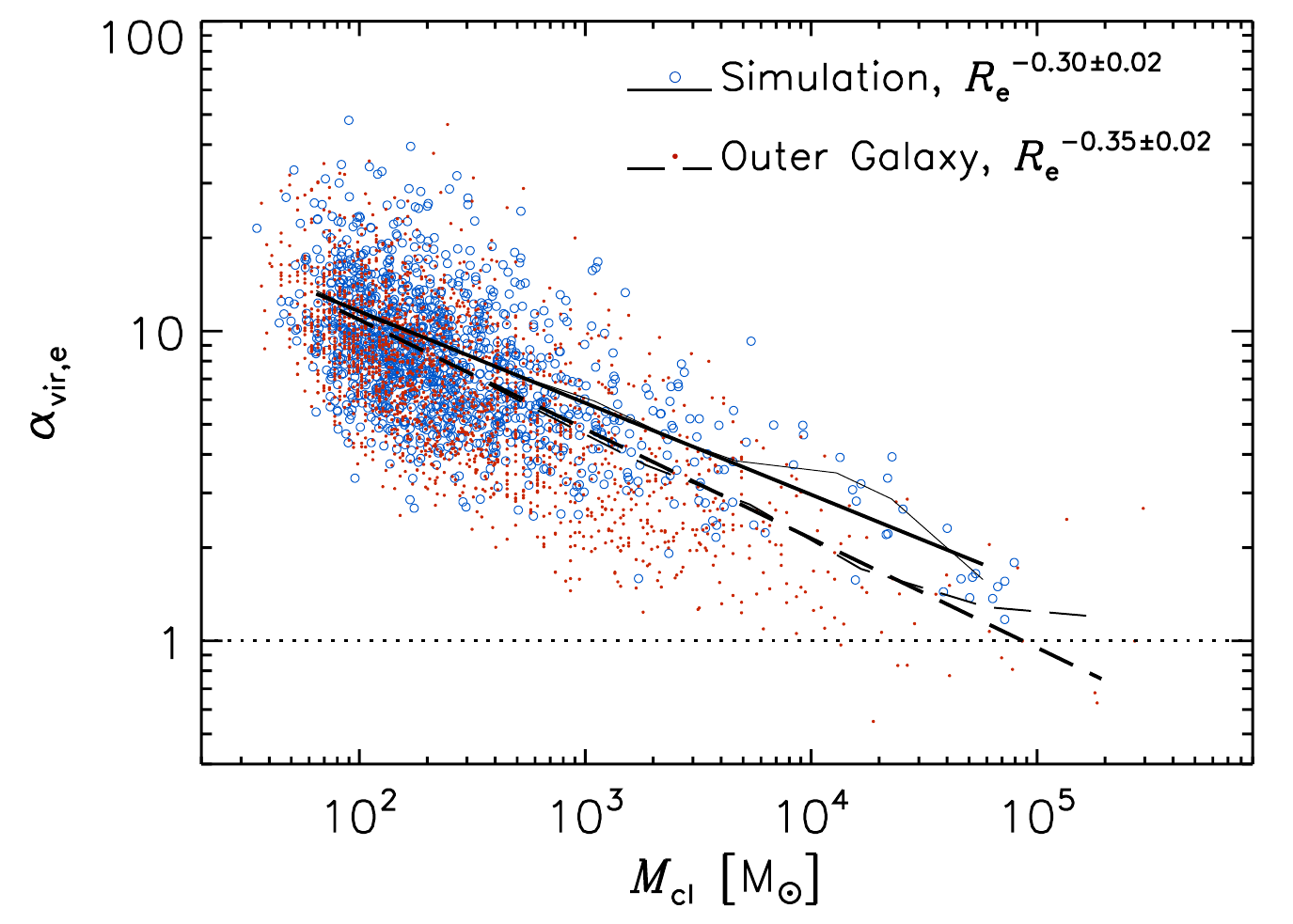}
\caption[]{Cloud virial parameter versus cloud mass for the same synthetic and observational samples as in the previous figures. The thin lines
are the averaged values of $\alpha_{\rm vir,e}$ in logarithmic intervals of cloud mass, and the thick lines their power-law fits. The lack of clouds
of small mass and small $\alpha_{\rm vir,e}$ is a selection effect due to the limited velocity resolution.}
\label{alpha}
\end{figure}

In Figure \ref{mass_size} we plot the mass-size relations. As found in previous observational studies \citep[e.g.][]{Roman-Duval+10}, the cloud surface 
density increases with increasing cloud size, in agreement with estimates of MC fractal dimensions from various observational surveys 
\citep[e.g.][]{Bazell+Desert88,Zimmermann+Stutzki92,Elmegreen+Falgarone96,Sanchez+07} and from simulations of randomly driven turbulence 
\citep[e.g.][]{Kritsuk+07,Federrath+09}. This trend is reproduced by our fiducial synthetic catalog, though with a slightly shallower slope than in the 
Outer Galaxy Survey. We follow the same binning procedure as for the velocity-size relation to obtain power-law fits, 
\begin{equation}
M_{\rm cl}=M_{\rm cl,1pc}\left({R_{\rm e}\over{1\,{\rm pc}}}\right)^{\beta_{\rm m}}
\label{eq_ms}
\end{equation}
with $M_{\rm cl,1pc}=65\pm 2$ M$_{\odot}$ and $\beta_{\rm m}=2.24\pm 0.04$ for the synthetic clouds, and $M_{\rm cl,1pc}=49\pm 2$ M$_{\odot}$  
and $\beta_{\rm m}=2.39\pm 0.05$ for the observed clouds. Both mass-size relations correspond to a mean cloud surface density of approximately
32 M$_{\odot}$\, pc$^{-2}$ at $R_{\rm e}= 6.5$ pc.

The information from the two Larson relations can be combined into a relation between the cloud virial parameter and the cloud mass, which reveals
the dynamical state of the MCs. The virial parameter is defined as
\begin{equation}
\alpha_{\rm vir,e} \equiv {5 \sigma^2_{\rm v} R_{\rm e} \over{G M_{\rm cl}}} \sim {2 E_{\rm k}\over{E_{\rm g}}},
\label{eq_alpha}
\end{equation}
where $\sigma_{\rm v}$ is the one-dimensional velocity dispersion, $G$ the gravitational constant, $E_{\rm k}$ the internal kinetic energy and 
$E_{\rm g}$ the gravitational energy, and the second equality is exact in the case of an idealized spherical cloud of 
uniform density \citep{Bertoldi+McKee92}. For more realistic cloud mass distributions, the virial parameter is only an approximation of the ratio of 
kinetic and gravitational energies. Because it adopts the equivalent radius, $R_{\rm e}$, we refer to this virial parameter as $\alpha_{\rm vir,e}$.
The dependence of $\alpha_{\rm vir,e}$ on cloud mass is shown in Figure \ref{alpha}, together with the power-law fits derived from binning the data 
in logarithmic intervals of $M_{\rm cl}$. Both the scatter and the trend of $\alpha_{\rm vir,e}$ versus $M_{\rm cl}$ of the synthetic clouds are approximately
consistent with the observational data. The exponents of the power-law fits are consistent with those derived from combining the Larson relations
expressed by equations (\ref{eq_larson}) and (\ref{eq_ms}), giving $(1+2\beta_{\rm v}-\beta_{\rm m})/\beta_{\rm m}$.
Notice that, while the upper envelope of the scatter plots in Figure \ref{alpha} is real, the lower envelope is a selection effect due to the limited
velocity resolution. With higher velocity resolution, low-mass clouds with low values of $\alpha_{\rm vir,e}$ would also be selected. 

Figure \ref{alpha} shows that the most massive clouds are more likely to be gravitationally bound than smaller clouds. Given the rather large values
of $\alpha_{\rm vir,e}$, the great majority of clouds must be transient structures. This result depends to some extent on the cloud definition 
as over-densities above a {\it fixed} brightness-temperature threshold, resulting in increasing surface density with increasing mass. One could always 
define clouds with a different criterion, for example {\it imposing} $\alpha_{\rm vir,e}=1$, which could be achieved by adopting a variable 
brightness-temperature threshold (increasing towards smaller masses), or with a dendrogram analysis \citep[e.g.][]{Rosolowsky+08}. However,
most reasonable definitions would lead to the conclusion that a majority of MCs from our simulation and from the Outer Galaxy survey are
transient structures.

\section{Spatial resolution and CO abundance} \label{sect_resol}

In the previous sections, we have compared our fiducial model, V2 at 0.98 pc resolution, with the subset of the observational 
sample at the corresponding distance of approximately 4 kpc. We now consider the effect of changing the CO abundance model 
and the spatial resolution of the synthetic observations. We maintain the velocity resolution fixed, to match that of the observations,
because our main focus is the comparison of the simulation with the Outer Galaxy Survey. We will study the effect of varying both
the spatial and the velocity resolution in a separate work. 

To look for possible trends related to the spatial resolution also in the observations, 
we divide the observed clouds into six slightly overlapping distance intervals, corresponding to the spatial resolution intervals 
$[x_i/1.25,x_i\times1.25]$, with $x_i= 0.5\times2^{i/2}$~pc, $i=0,... 5$ (see Table \ref{t1}). The number
of observed clouds, $N_i$, in the $x_i$ intervals are $N_{0,... 5}=212, 584, 1936, 1816, 360, 88$. The third interval is the most
populated one and has an average resolution of 1.04 pc, nearly identical to that of our fiducial model. This procedure of dividing 
the observed clouds into distance intervals is not exactly the same as re-observing with different resolutions clouds at a fixed distance,
as in the synthetic observations. In the real observations, clouds are selected from maps containing objects with a rather large range of 
distances. Thus, the following comparison of the dependence on resolution of simulated and observed clouds should be considered 
of a qualitative nature. 

\begin{figure}[t]
\includegraphics[width=\columnwidth]{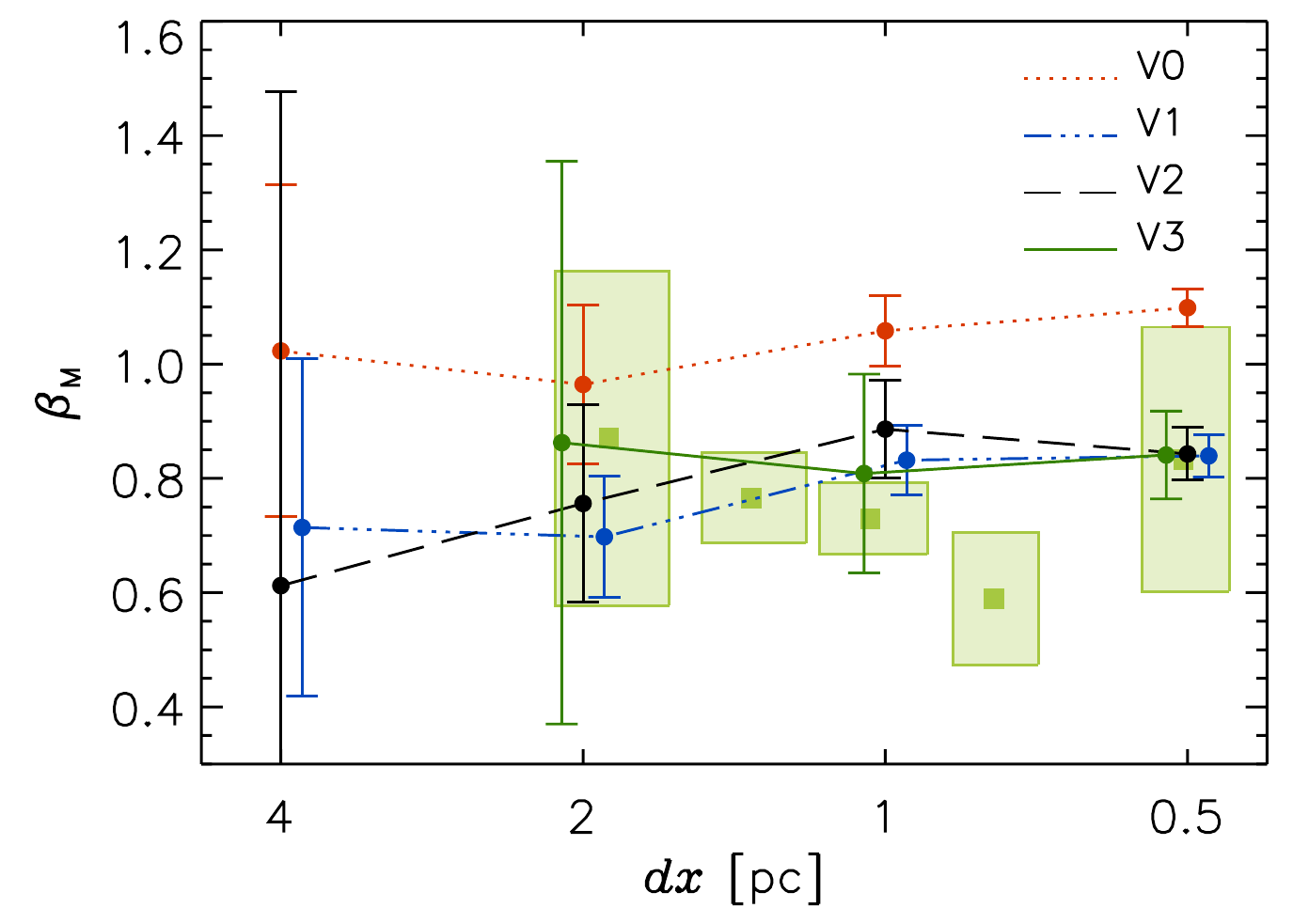}
\caption[]{Power-law exponents of the fits to the mass distributions versus spatial resolution. Each plot shows a different model for the assumed relative 
abundance of CO, with the error bars giving the 1-$\sigma$ uncertainty of the estimated slope. The squared symbols are the values from the different 
distance intervals of the observational sample (see text) and the shaded areas show the standard deviation of the mean spatial resolution within each 
interval and the estimated 1-$\sigma$ uncertainty of $\beta_{\rm M}$. The lowest-resolution interval of the observations and the lowest-resolution value
of model V3 are not shown due to their excessive uncertainty.}
\label{ex_mass}
\end{figure}

\begin{figure}[t]
\includegraphics[width=\columnwidth]{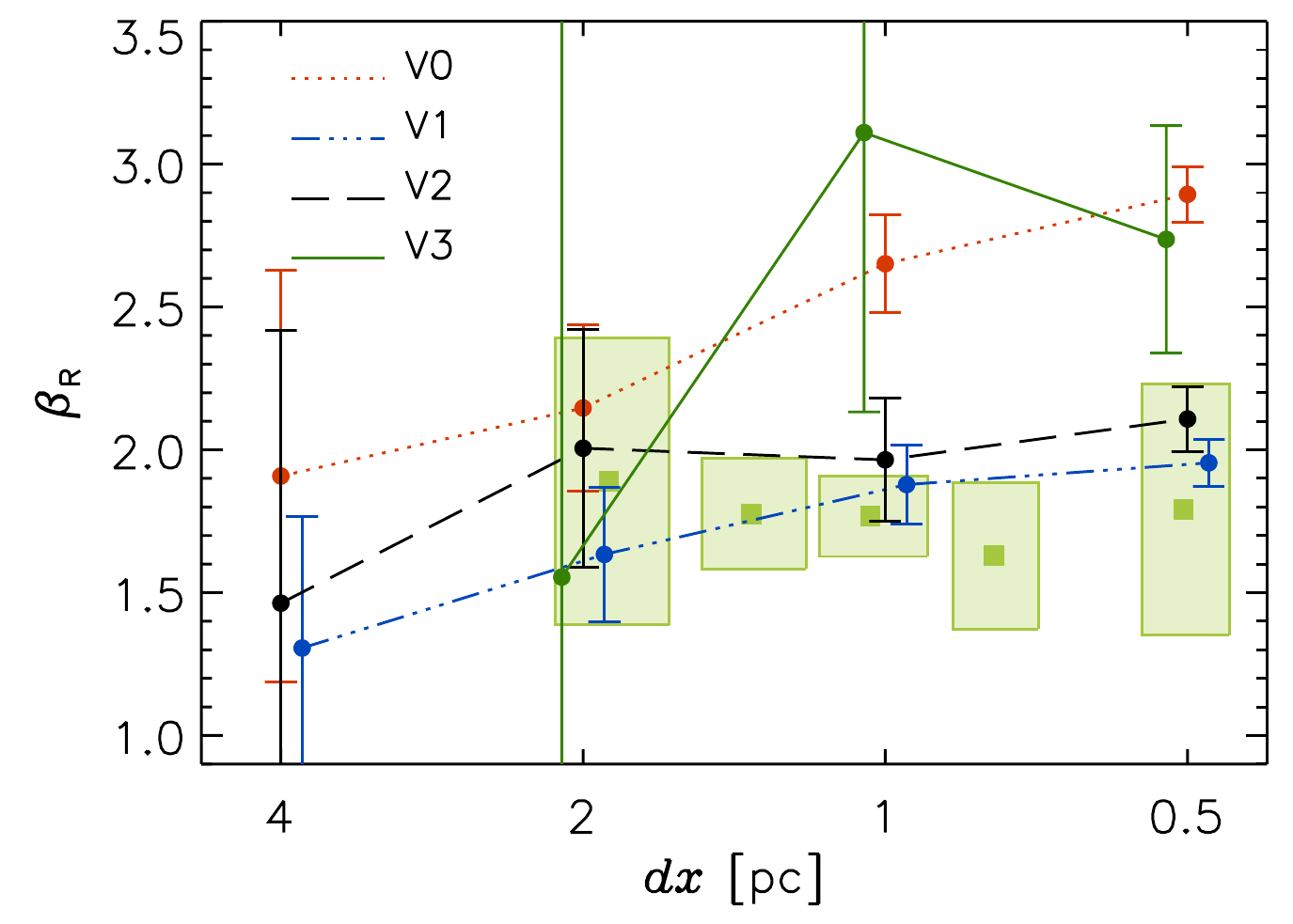}
\caption[]{Power-law exponents of the fits to the size distributions versus spatial resolution. As in the previous figure, each plot shows a different model 
for the assumed relative abundance of CO, with the error bars giving the 1-$\sigma$ uncertainty of the estimated slope. The squared symbols are the 
values from the different distance intervals of the observational sample (see text) and the shaded areas show the standard deviation of the mean spatial 
resolution within each interval and the estimated 1-$\sigma$ uncertainty of $\beta_{\rm R}$.}
\label{ex_size}
\end{figure}

We first consider the dependence of the slopes of the mass and size distributions, $\beta_{\rm M}$ and $\beta_{\rm R}$
respectively, on the CO abundance model and on the resolution, shown in Figures \ref{ex_mass} and \ref{ex_size}. The 
squares show the mean spatial resolution and mean slope derived from the observations in each distance interval (the 
lowest-resolution bin is not shown because the power-law fit to the histograms has too large uncertainty due to the small 
number of clouds). The shaded areas give the 1-$\sigma$ uncertainty in the estimated slope and the standard deviation 
of the spatial resolution values (the statistical uncertainty in the resolution, from the cloud distance uncertainty, may be 
even larger).


The power-law fits are derived for masses and radii above some minimum values that are set to be the same for the synthetic and the observational 
MC catalogs, and just above the estimated mass and size completeness limits, as mentioned in section \ref{sect_mass_size}. For the 1 pc resolution, 
the power-law fits are computed for masses larger than 200 M$_{\odot}$ and radii larger than 2 pc. For different resolutions, the minimum mass is 
scaled as $dx^2$ and the minimum radius as $dx$.

Figures \ref{ex_mass} and \ref{ex_size} show that our fiducial chemical abundance model, V2, yields mass and size distributions with 
slopes approximately independent of spatial resolution, within the 1-$\sigma$ uncertainties. The same is true for the observations (shaded areas) 
and for the other chemical models, except for a significant increase in $\beta_{\rm R}$ between $dx=2$ and 0.5 pc in the V0 model. Furthermore, 
the slopes derived from the synthetic observations with variable CO abundance models are also in good agreement with the slopes from the observations. 
On the contrary, the slopes from the constant CO abundance model, V0, are significantly larger than the observational slopes. The assumption of constant 
CO abundance results in the selection of a large number of small clouds, as illustrated by the red contours in Figure \ref{images}, causing the steeper 
mass and size distributions. 

The values of $\beta_{\rm M}$ and $\beta_{\rm R}$ from the model V1 are indistinguishable from those of model V2 and thus
consistent with the observations as well. They also have smaller error bars than in model V2 due to the larger number of clouds (see Table \ref{t1}).
Finally, model V3 is also consistent with models V1 and V2 and with the observations, but its error bars are very large due to the small sample size.

\begin{figure}[t]
\includegraphics[width=\columnwidth]{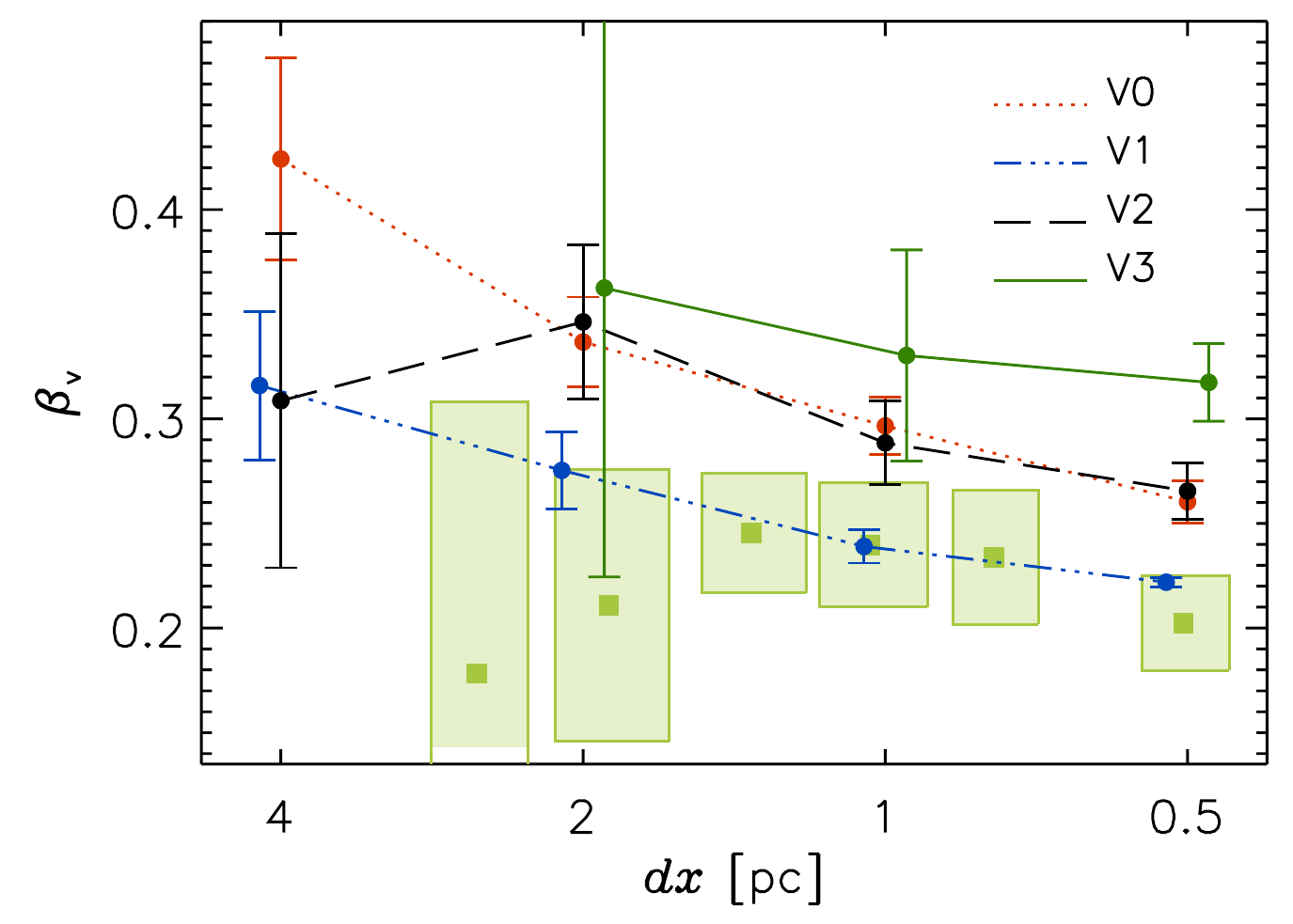}
\caption[]{As in Figures \ref{ex_mass} and \ref{ex_size}, but for the exponents of the velocity-size relation. All six cloud distance intervals
of the observational samples are shown here.}
\label{ex_velocity}
\end{figure}

Figure \ref{ex_velocity} shows the values of the exponent of the velocity-size relation, $\beta_{\rm v}$, derived from power-law fits of the data
binned in logarithmic size interval, as in the previous section. All models yield a slight decrease in $\beta_{\rm v}$ with increasing resolution,
at least for $dx \le 2$ pc. A similar trend is also visible in the values from the observations, but it is not significant within the 1-$\sigma$ uncertainty.
The V1 model provides the best match to the observations. The model V2 is also consistent with the observations at the 1-$\sigma$ level at our
fiducial resolution of 0.98 pc, while it gives slightly too large values at $dx=0.5$ and 2. The model V0 never matches the observations. Its values of
$\beta_{\rm v}$ are very close to those of model V2, but its error bars significantly smaller, due to the larger sample size, so its discrepancy with 
the observations is significant. In the case of model V3 the error bars are very large, but the discrepancy with the observations is nevertheless 
very large at $dx =0.5$ pc.

\begin{figure}[t]
\includegraphics[width=\columnwidth]{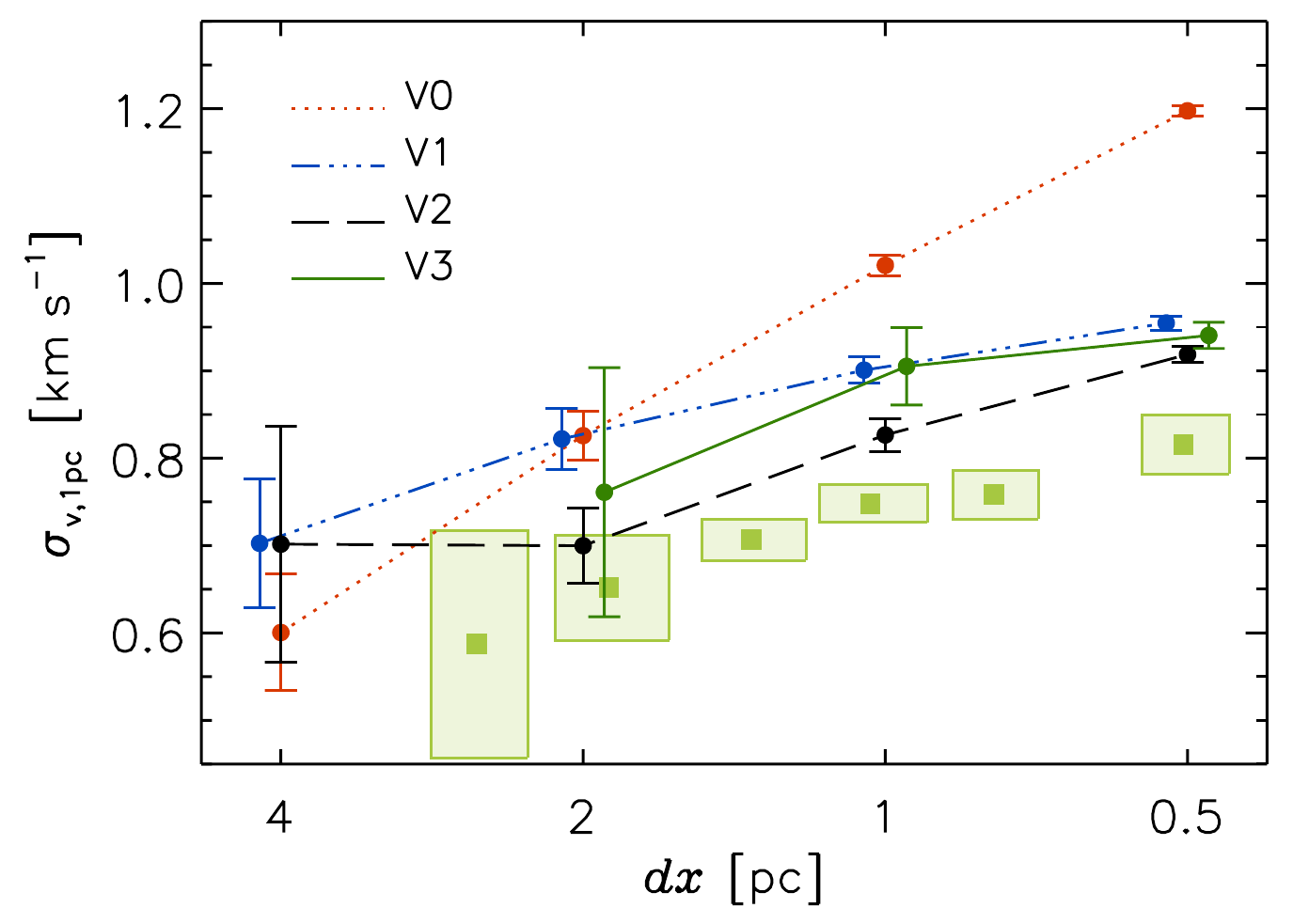}
\caption[]{As in Figure \ref{ex_velocity}, but for the normalizations coefficient of the velocity-size relation.}
\label{norm_velocity}
\end{figure}

The normalization coefficient of the velocity-size relation, $\sigma_{\rm v,1pc}$, has a much stronger dependence on spatial resolution than $\beta_{\rm v}$. 
In Figure \ref{norm_velocity}, all models show an almost linear increase of the rms velocity at the scale of 1 pc, $\sigma_{\rm v,1pc}$, 
with the logarithm of the resolution. The observational catalogs also follow the same trend, with $\sigma_{\rm v,1pc}\approx 0.58$ km\,s$^{-1}$ at 
$dx\approx 2.55$ pc, and $\sigma_{\rm v,1pc}\approx 0.82$ km\,s$^{-1}$ at $dx\approx 0.50$ pc. This effect is probably due to the limited velocity 
resolution of approximately 0.5 km\,s$^{-1}$. When the spatial resolution is increased, some of the clouds are broken into smaller-size components, 
while this further fragmentation is not matched in velocity space, where the resolution is held constant. Based on the dependence of the cloud sample 
size on resolution in the synthetic catalogs, we can estimate that the characteristic cloud size decreases almost linearly with $dx$. Assuming that the 
cloud line width is nearly unchanged by the fragmentation on the plane of the sky, we would expect that $\sigma_{\rm v,1pc}$ grows with decreasing 
$dx$ with a slope not much smaller than $\beta_{\rm v}$. That is indeed the case for both the observations and all the synthetic models with variable 
CO abundance. On the other hand, the V0 model has a significantly steeper dependence on resolution. That may be due to the combined effect
of the steeper mass and size distributions, yielding a much larger number of small clouds, and the inclusion of low-density high-velocity gas that is not 
probed by the variable-abundance models.

\begin{figure}[t]
\includegraphics[width=\columnwidth]{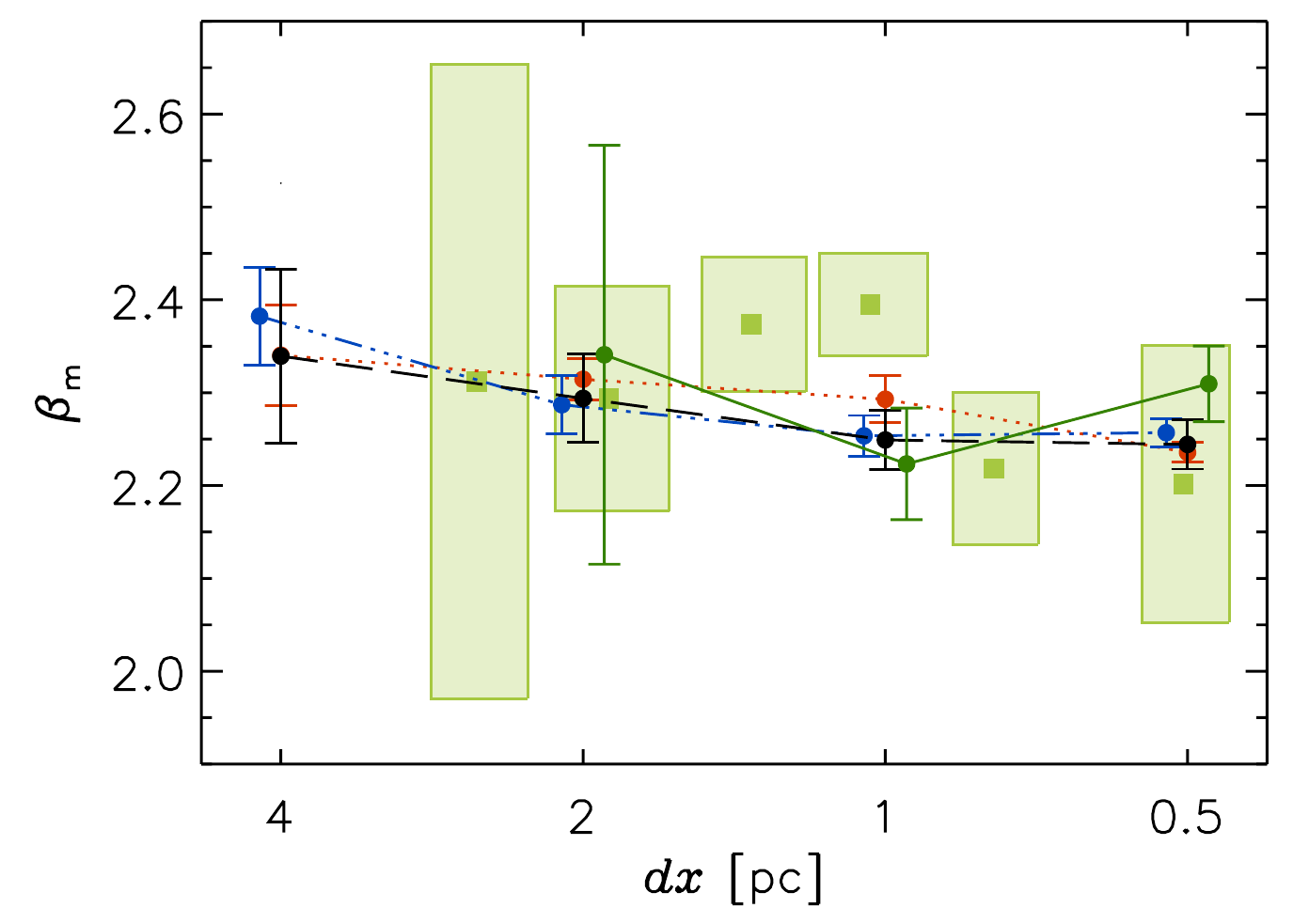}
\caption[]{Exponents of the fits to the mass-size relation, derived as for the velocity-size relation of Figure \ref{ex_velocity}.}
\label{ex_ms}
\end{figure}

Although they follow the same trend with resolution as the observations, the models have generally too large values of $\sigma_{\rm v,1pc}$
relative to the observations. As commented in the previous section, this is to be expected because the SN rate in the simulation is not  
tailored to match that in the Perseus arm and it is relatively large. We argued above that our fiducial model, V2 at 0.98 pc resolution, would
yield a value of $\sigma_{\rm v,1pc}$ consistent with the observations if the SN rate were reduced to 70\% of its current value. The
correction factor is approximately the same also at $dx=0.5$ and 2 pc, although the linear scale in $\sigma_{\rm v,1pc}$ and the decreasing 
error bar with decreasing $dx$ in Figure \ref{norm_velocity} give the appearance of a larger discrepancy at higher resolution. 
The normalization of the velocity-size relation is significantly larger in the other two variable-abundance models, V1 and V3, and much larger in
model V0 at $dx=0.5$ and 1 pc, due to the stronger dependence on resolution of this model. Furthermore, the dependence on resolution in
model V1 is significantly weaker than in model V2 and in the observations. In model V1, the SN rate should be reduced by a factor 1.55 
at $dx=0.5$ pc and 2.5 at $dx=2$ pc in order to match the observed values of $\sigma_{\rm v,1pc}$.

\begin{figure}[t]
\includegraphics[width=\columnwidth]{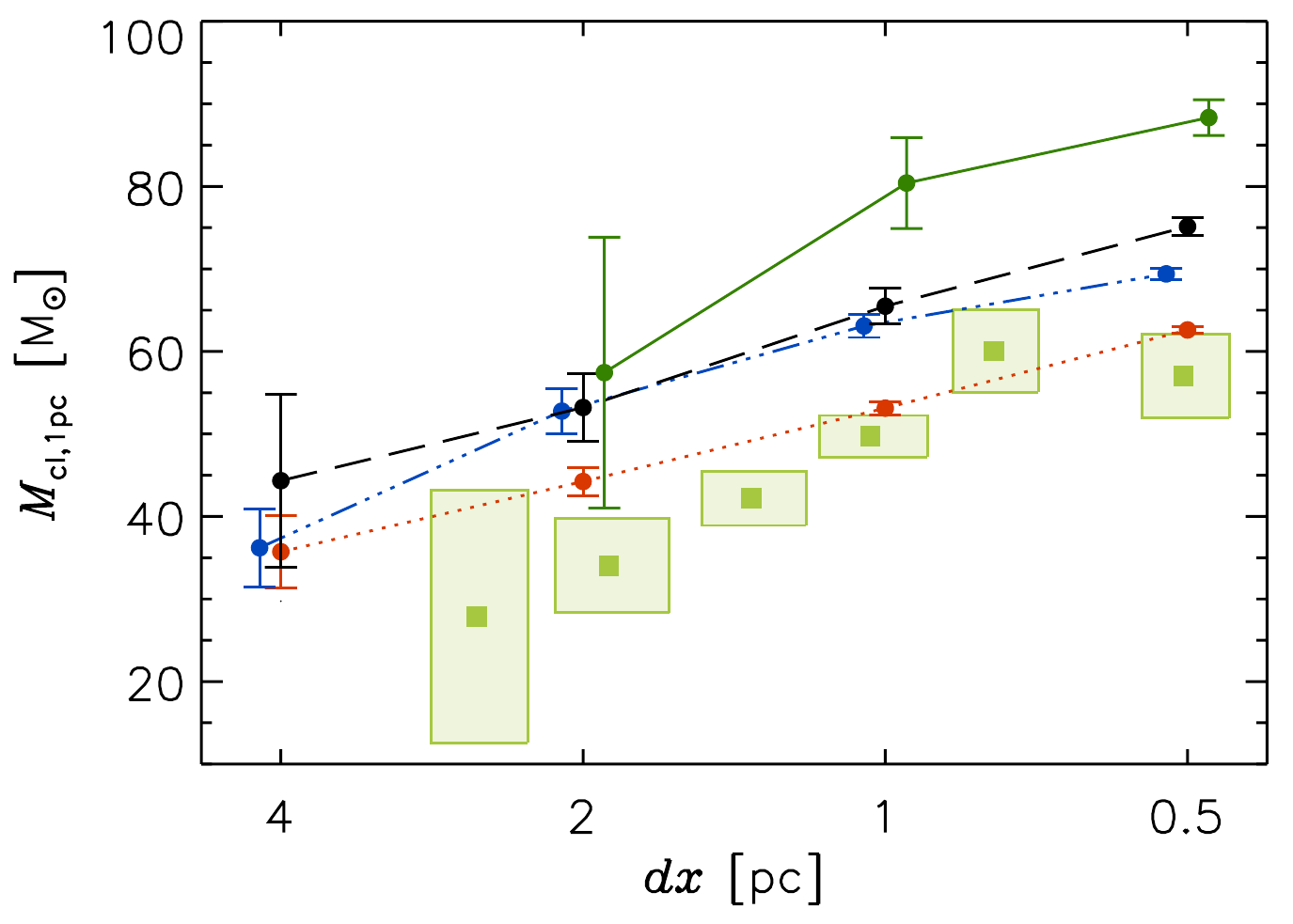}
\caption[]{Normalizations coefficients of the mass-size relation. Same symbols as in the previous figures.}
\label{norm_ms}
\end{figure}

The dependence of the mass-size relation on the CO abundance model and spatial resolution is shown in Figures \ref{ex_ms} and \ref{norm_ms}.
In the synthetic catalogs, the exponent of the power law fits, $\beta_{\rm m}$, is essentially independent of both resolution and CO abundance. The
same is true for the observations, although there is a significant drop as $dx$ decreases from 1 pc to 0.8 pc. The values from the synthetic catalogs 
are consistent with the observations, except around $dx=1$ pc, where the mass-size relation from the observations is a bit steeper than in the models, 
as already anticipated in Figure \ref{mass_size}. As in the velocity-size relation, the normalization coefficient of the mass-size relation has a clear dependence 
on resolution. $M_{\rm cl,1pc}$ increases with decreasing $dx$ in all the synthetic catalogs and in the observations. The only exception is the highest resolution 
interval of the observational sample. In all models but V0 the clouds are a bit more massive than the observations at $R_{\rm e}=1$ pc. This discrepancy is partly 
due to the steeper slope of the observational mass-size relation, causing the discrepancy in the normalization at 1 pc, while the cloud masses are comparable at 
intermediate cloud sizes of a few pc. As shown in Table 1, the mean MC surface density is $\approx 20 M_{\odot}$~pc$^{-2}$ for both the synthetic and observational
samples, with only a slight increase in surface density as the CO abundance is reduced from model V0 to model V3. However, because the simulation is not specifically
tailored to match MCs in the Perseus arm, we should not expect the synthetic catalogs to yield exactly the same surface density as the observations.

Based on this discussion of the dependence on CO abundance model and resolution, we can conclude that the constant abundance model, V0,
is mostly inconsistent with the observations, as its mass and size distributions and its velocity-size relation are too steep at all resolutions. Furthermore,
the normalization of its velocity-size relation is both too large and too strongly dependent on spatial resolution. Matching the model V0 values of 
$\sigma_{\rm v,1pc}$ with the observational ones would require a reduction of the SN rate in the simulation by a factor of 3.1 at $dx=2$ pc and
a factor of 2.6 at $dx=0.5$ pc. The inconsistency of model V0 with the observations was to be expected, as the assumption of constant CO
abundance is not realistic, and this model was included only as a reference. 

Among the variable-abundance models, even allowing for its large error bars, model V3 provides an inferior match to the observations, relative to models V1 
and V2, both with respect to $\beta_{\rm R}$ (too steep size distribution at $dx \ge 1$ pc) and $\beta_{\rm v}$ (too steep velocity-size relation at $dx=0.5$ pc).
Model V1 provides a better fit to the slope of the observed velocity-size relation than model V2. On the other hand, the dependence on resolution of 
$\sigma_{\rm v,1pc}$ in model V1 is weaker than in the observations, while in model V2 it is consistent with the observations. In summary,
both models V1 and V2 compare reasonably well with the observations. While model V2 may be considered more realistic than model V1, as it 
represents the average relation between CO abundance and gas density from a realistic simulation including the evolution of a chemical network 
\citep{Glover+Clark12}, it neglects the large scatter in such relation found in the chemistry simulations.

\section{Summary and Conclusions} \label{sect_conclusions}

We have carried out a direct comparison of MCs selected from a simulation of SN-driven ISM turbulence with MCs selected from the Outer Galaxy Survey,
focusing on the mass and size distributions and on the Larson relations. A similar comparison was presented in Paper I, but only for clouds
selected in PPP space and without any attempt to model the relative abundance of CO or the radiative transfer of CO emission lines. Here, we have used three
snapshots of the simulation to compute synthetic CO observations with the same spatial and velocity resolutions and the same noise level as in the Outer 
Galaxy Survey, and we have selected the clouds following closely the procedure described in \citet{Heyer+01}. As in Paper I, we find a
reasonable agreement between the simulation and the observations, suggesting that SN-driven turbulence may be the main agent responsible for the 
formation and evolution of MCs. However, we have also shown that our results are sensitive to the assumptions about the CO abundance. We have not 
modeled the ISM chemical evolution, and have relied on average relations between CO abundance and gas density from previous works, neglecting 
the large scatter found in those relations. Future tests of ISM turbulence simulations based on the comparison with observed MC properties will have to include
self-consistently the solution of chemical networks to compute the spatial distribution of the observed molecular species.

The main results of this work are summarized in the following.

\begin{enumerate}

\item The mass and size distributions from the synthetic MC catalogs with variable CO abundance are consistent with the observations. Their slopes do not
vary significantly with spatial resolution nor with chemical model. The synthetic MC catalog with constant CO abundance yields too steep mass and size distributions.

\item The mean surface density of the synthetic clouds is $\approx 20 M_{\odot}$~pc$^{-2}$, comparable to that of the MCs from the Outer Galaxy survey. 

\item The velocity-size relation is slightly steeper in the synthetic MCs than in the observations, except for the case of the chemical abundance model V1. 
Model V2 is also consistent with the observations in the resolution interval around 1 pc, within the 1-$\sigma$ uncertainty. The exponent of the velocity-size
relation has a slight dependence on resolution.   

\item The normalization of the velocity-size relation has a clear dependence on resolution, both in the synthetic catalogs and in the observations. As the spatial 
resolution is increased, while keeping the velocity resolution constant, the mean velocity and 1 pc, $\sigma_{\rm v,1pc}$, increases. This behavior of 
$\sigma_{\rm v,1pc}$ is to be expected, as smaller-size clouds are selected at higher spatial resolution without a significant decrease in their line width due to the
constant velocity resolution, but the effect had never been demonstrated with real or synthetic observations prior to this work. This dependence of $\sigma_{\rm v,1pc}$ on
resolution is much too steep in the constant-abundance V0 model compared with the observations. 

\item The normalization of the mass-size relation at 1 pc, $M_{\rm cl,1pc}$, depends on spatial resolution as well, both in the synthetic catalogs and in the observations.
It is a bit higher in the synthetic catalogs, particularly at the resolution $dx=1$ pc where the observational mass-size relation is significantly steeper than in the simulation.

\item The normalization of the velocity-size relation of all the synthetic catalogs is larger than in the observations. In the case of our fiducial model, this suggests that 
the SN rate in the Perseus arm may be approximately 70\% of the value used in the simulation. 

\end{enumerate}

The subdivision of the observational sample in cloud distance intervals is not exactly equivalent to the procedure of generating synthetic catalogs of different
spatial resolutions. Thus, the comparison of the synthetic catalogs with the observations based on the dependence on spatial resolution should be considered 
to be of a qualitative nature. Focusing on the results at 1 pc resolution, where we have the largest number of observed clouds, we can still conclude that the 
constant-abundance model is ruled out, while all variable abundance models are consistent with the observations (a bit beyond the 1-$\sigma$ uncertainty in the 
case of model V3 that has the smallest sample size and the largest error bars), except for a slightly too large average velocity dispersion and too shallow mass-size
relation (causing a too large value of $M_{\rm cl,1pc}$, although the average surface density matches the observations nicely). Although
the velocity dispersion is proportional to the cubic root of the SN rate, it appears that the velocity-size relation of MCs selected with a constant
brightness temperature threshold provides a rather sensitive constraint for the SN rate.

\acknowledgements

We thank the referee for useful comments that helped us improve the manuscript.
Computing resources for this work were provided by the NASA High-End Computing (HEC) Program through 
the NASA Advanced Supercomputing (NAS) Division at Ames Research Center, by PRACE through a Tier-0 award providing 
us access to the computing resource SuperMUC based in Germany at the Leibniz Supercomputing Center, and by the Port 
d'Informaci\'{o} Cient\'{i}fica (PIC), Spain, maintained by a collaboration of the Institut de F\'{i}sica d'Altes Energies (IFAE) 
and the Centro de Investigaciones Energ\'{e}ticas, Medioambientales y Tecnol\'{o}gicas (CIEMAT). PP acknowledges support 
by the ERC FP7-PEOPLE- 2010- RG grant PIRG07-GA-2010-261359 and by the Spanish MINECO under project AYA2014-57134-P. 
MJ acknowledges support by the Academy of Finland grant 1285769.
TH is supported by a Sapere Aude Starting Grant from The Danish Council for Independent Research. 
Research at Centre for Star and Planet Formation was funded by the Danish National Research Foundation and the University
of Copenhagen's programme of excellence.


\end{document}